# High-Temperature Refractory Metasurfaces for Solar Thermophotovoltaic Energy Harvesting

Chun-Chieh Chang[†], Wilton J. M. Kort-Kamp[‡,§], John Nogan[∥], Ting S. Luk[∥], Abul K. Azad[†], Antoinette J. Taylor[⊥], Diego A. R. Dalvit[§], Milan Sykora[#], and Hou-Tong Chen[*,†]

[†]Center for Integrated Nanotechnologies, Los Alamos National Laboratory, Los Alamos, New Mexico 87545, USA

[‡]Center for Nonlinear Studies, Los Alamos National Laboratory, Los Alamos, New Mexico 87545, USA

[§]Theoretical Division, Los Alamos National Laboratory, Los Alamos, New Mexico 87545, USA

[∥]Center for Integrated Nanotechnologies, Sandia National Laboratories, Albuquerque, New Mexico, 87123, USA

[⊥]Chemistry, Life, and Earth Sciences Directorate, Los Alamos National Laboratory, Los Alamos, New Mexico 87545, USA

[#]Chemistry Division, Los Alamos National Laboratory, Los Alamos, New Mexico 87545, USA

KEYWORDS: Metasurfaces, refractory metamaterials, solar thermophotovoltaics, solar absorbers, thermal emitters, high-temperature

**ABSTRACT**

Solar energy promises a viable solution to meet the ever-increasing power demand by providing a clean, renewable energy alternative to fossil fuels. For solar thermophotovoltaics (STPV), high-temperature absorbers and emitters with strong spectral selectivity are imperative to efficiently couple solar radiation into photovoltaic cells. Here, we demonstrate refractory metasurfaces for STPV with tailored absorptance and emittance characterized by *in-situ* high-temperature measurements, featuring thermal stability up to at least 1200 ºC. Our tungsten-based metasurface absorbers have close-to-unity absorption from visible to near infrared and strongly suppressed emission at longer wavelengths, while our metasurface emitters provide wavelength-selective emission spectrally matched to the band-edge of InGaAsSb photovoltaic cells. The projected overall STPV efficiency is as high as 18% when employing a fully integrated absorber/emitter metasurface structure, much higher than those achievable by stand-alone PV cells. Our work opens a path forward for high-performance STPV systems based on refractory metasurface structures.

**TEXT**

Photovoltaics (PV)[1] directly convert sunlight to electricity using semiconductor PV cells, and have been the most prevalent solar energy-harvesting technology. Despite the development over the past few decades, the efficiency of state-of-the-art, single-junction PV cells is still far below the fundamental limit predicted by Shockley and Queisser[2], which is dictated mainly by energy losses due to below-bandgap photons and hot-carrier thermalization, owing to the broad distribution of the solar spectrum. To minimize these losses, numerous novel PV device concepts have been proposed and realized[3-7]. While they indeed improve the PV efficiency to some extent, they all suffer from their own respective problems, including high manufacturing cost, complex device fabrication processes, as well as material instability and degradation. Solar thermophotovoltaics (STPV)[8, 9] represent a promising alternative to traditional photovoltaics for solar energy harvesting, where an absorber/emitter intermediate structure first absorbs the incoming sunlight, heats up, and then emits thermal photons towards the PV cell to excite charge carriers for power generation. An ideal STPV system has a solar-to-electric energy conversion efficiency much higher than that of a stand-alone PV cell, as a carefully designed STPV intermediate structure can fully capture the incident sunlight and convert it into narrowband thermal emission right above the bandgap of the PV cell[10]. It has been theoretically shown that the STPV efficiency could significantly surpass the aforementioned Shockley-Queisser limit, reaching 85% and 54% under fully concentrated and unconcentrated solar radiation, respectively[11].

Recently, several proof-of-concept STPV experiments have been reported employing various absorber/emitter intermediate structures[12-16], including multi-walled carbon nanotubes, photonic crystals (PhCs), and two-dimensional multilayers. Although these initial demonstrations are quite encouraging, the realized energy conversion efficiency is, however, limited to a few percent, and significant work remains to further improve the spectral selectivity and high-temperature stability of absorbers and emitters for enhanced STPV efficiencies. Solar absorbers for STPV should have high absorptance in the visible and near-infrared spectrum but strongly suppressed mid-infrared re-emission, and should resist degradation at elevated temperatures (>1000 °C). Solar absorbers can be readily realized using transition metals and heavily doped semiconductors[17], tandem structures based on metal/dielectric multilayers[18, 19], and engineered micro- and nano-optical materials[20, 21]. However, these approaches suffer from either poor spectral selectivity, complex fabrication processing or, with a few exceptions[14, 22, 23], operation only at low temperatures (constrained by the melting point of their constituent materials, e.g., Au and Ag), and hence are not of practical use for STPV applications. Development of wavelength-selective thermal emitters for STPV can be even more challenging, as all materials at elevated temperatures naturally emit a fairly broad thermal radiation background[24]. To maximize the radiative heat transfer, the thermal emission should have close-to-unity emittance at energies right above the bandgap of the PV cell and spectrally alignment to the blackbody spectral maximum at the operational temperature. For PV cells fabricated from GaSb[25] or InGaAsSb[26] (bandgap energy 0.7 eV and 0.55 eV, respectively), these conditions require operational temperatures exceeding 1000 °C. To date, several selective thermal emitters intended for STPV have been demonstrated, employing metal gratings[27, 28], bull's eye structures[29], metallic PhCs[30-33], thin-film stacks[15, 34], and metamaterials[35-40]. However, except for a very few demonstrations[15, 33, 41], they either do not show thermal stability above 1000 °C or operate at energies much lower than the bandgap of the PV cells. Hence, there remains an urgent need for high-performance, thermally stable solar absorbers and thermal emitters with optimized spectral selectivity for STPV applications.

Here, we experimentally demonstrate refractory tungsten (W) metasurface solar absorbers and thermal emitters for STPV intermediate structures. The metasurface structures exhibit desirable wavelength selective absorptance and emittance, and their structural integrity does not show any significant degradation following exposure to repeated heating cycles between room temperature and 1200 °C. Furthermore, the absorptance/emittance spectra are thermally stable up to at least 1200 °C in vacuum by *in situ* high temperature measurements. Using a detailed balance calculation, we predict that a fully integrated intermediate structure based on our metasurface absorber/emitter would yield an STPV efficiency of 18% under 4000 suns concentrated solar radiation and operated at 1380 °C, much higher than that of a stand-alone InGaAsSb PV cell.

**Results**

The design of the refractory solar absorbers is based on a metal/dielectric/metal metasurface structure[42-44], using tungsten (W) for the metal and aluminum oxide ($Al_2O_3$) for the dielectric spacer owing to their high melting points (> 2000 °C) and low thermal expansion coefficients. As shown in Figure 1a, a unit cell comprising four W-nanodisks of different diameters was employed for maximizing absorption bandwidth in the solar spectral region while minimizing the absorption in the mid-infrared, through systematically tuning the geometrical parameters of the unit cell in full-wave numerical simulations. The optimized W-metasurface solar absorbers were then fabricated using electron beam lithography (EBL) methods with DC sputtered tungsten and atomic layer deposited (ALD) $Al_2O_3$ films (see Methods), as illustrated in the top-view scanning electron microscopy (SEM) image in Figure 1b. The reflectance $R(\lambda, \theta, \varphi)$ was first experimentally measured at room temperature (see Methods), from which the absorptance $A(\lambda, \theta, \varphi)$ was obtained using $A(\lambda, \theta, \varphi) = 1 - R(\lambda, \theta, \varphi)$, as the transmission was completely blocked by the W-ground plane. As shown in Figure 1c, the measured absorptance is higher than 90% within the entire visible and part of the near-infrared region where the solar spectrum has significant energy intensity (see the AM 1.5 solar spectrum indicated by the shaded background in Figure 1c). Towards longer wavelengths the absorption drops significantly: ~40% at $\lambda = 1.5$ μm, and well below 20% in the mid-infrared region. The measured absorptance is in overall agreement with the simulated spectra also shown in Figure 1c. As compared to bare W-films, the absorptance of the W-metasurface absorbers is significantly enhanced near visible wavelengths as shown by the black color of the optical micrograph for a fabricated sample in the inset to Figure 1b. At longer wavelengths the absorptance is suppressed and approaches that of a bare W film in the mid-infrared. The measured absorptance exhibits a weak dependence on the angle of incidence $\theta$ (see Figure 1d) and the azimuthal angle $\varphi$ (not shown), and does not depend on the incident polarization states due to their rotational symmetry. These properties are crucial for the STPV intermediate structure operating at high solar concentrations.

We found that the fabrication process of the W thin films (i.e., sputtering vs. e-beam evaporation) has a significant impact on the performance of the metasurface absorbers as shown in Figure 1e. The metasurface absorber using e-beam evaporated W-films exhibits high absorptance in the visible and near-infrared regions, but the absorptance remains high at longer wavelengths, which is undesirable as it causes a higher radiation loss. The contrast of absorption spectra can be understood by the very different optical and structural properties of the W-films deposited by sputtering and evaporation (see Supplementary Figures S1 and S2). In order to eventually operate at high temperatures, we also deposited a conformal 20 nm-thick hafnium oxide ($HfO_2$) layer onto the fabricated W-metasurface absorbers as the protective layer to prevent the W-nanodisks from possible oxidation and evaporation and improve their thermal stability[29, 33, 45]. The absorptance spectrum measured at room temperature is shown in Figure 1f, exhibiting the same characteristic as that of the uncoated one, but slightly red-shifted. This is expected as the HfO$_2$ coating changes the dielectric environment of the resonant W-nanodisks, and can be easily compensated by tailoring their geometrical parameters such as the nanodisk diameters.

According to Kirchhoff's law, the design of a wavelength-selective thermal emitter is equivalent to that of a wavelength-selective absorber. Similar to the broadband metasurface absorbers described above, we employed single-size W-nanodisks to demonstrate our metasurface thermal emitters, as shown in Figure 2a for the unit cell schematic and the left inset to Figure 2d for an SEM image of a fabricated sample. The room-temperature emittance spectra $E(\lambda, \theta, \varphi)$ are shown in Figure 2b for two selected nanodisk diameters, obtained from the measured absorptance by using $E(\lambda, \theta, \varphi) = A(\lambda, \theta, \varphi)$. A high emittance peak of ~90% can be observed, which is red-shifted from $\lambda$ = 1.45 μm to 1.65 μm as expected when the nanodisk diameter increases from $d$ = 360 nm to 400 nm. As a comparison, the emittance of sputtered plain W-films, as shown in Figure 1c, is merely 25 ~ 30% in the same wavelength range. The wavelength selective emission in the near-infrared can be efficiently coupled into InGaAsSb PV cells[46, 47]. As shown in Figure 2b, the emittance peak of the $d$ = 360 nm emitter is located at the maximum external quantum efficiency (EQE) of InGaAsSb PV cells. We notice the broad emittance band near the visible region, with magnitude and spectral position almost independent on the nanodisk size. However, owing to the low spectral radiance of the blackbody at short wavelengths[24] for $T < 2000$ °C shown as the shaded background in Figure 2b, this high emittance in the visible region contributes very little to the emitted output power, and hence does not degrade the spectral selectivity. In Figure 2c we show the measured angular dependence of the emittance for the W-metasurface emitter with $d$ = 400 nm. It is apparent that the emittance peak at $\lambda$~1.65 μm does not shift as the angle increases, and remains about 60% when the angle increases to 60°, consistent with the typical behavior of metasurface absorbers[48]. This is in sharp contrast to photonic crystal thermal emitters where the emittance drops significantly at large angles. Therefore our metasurface emitter can provide much larger hemispherical integrated emission power, which is especially advantageous for STPV systems with a cylindrical geometry[49], as the thermal energy emitted in all directions from the emitter can be fully absorbed by its surrounding PV cells for power generation.

Practical metasurface intermediate structures for STPV should maintain their optical and structural properties after long-time operation and repeated heating cycles. The thermal stability of the W-metasurfaces was investigated using an in-house high-temperature optical characterization set-up (see Methods and Supplementary Figure S3) by applying multiple heating cycles between room temperature and up to 1200 °C (see Methods). The metasurface spectral absorptance/emittance were then characterized at room temperature and/or at 1200 °C, and their structural morphology was examined via SEM. In Figure 2b we show the room-temperature emittance measured before and after 10 heating cycles for the W-metasurface thermal emitters. Although they retain good structural integrity, as can be seen by the SEM image in the right inset to Figure 2d, the emittance peaks in the near-infrared are red-shifted by ~500 nm. For the $d$ = 360 nm (400 nm) emitter, its emittance peak shifts from $\lambda \sim$ 1.45 μm (1.65 μm) to $\lambda \sim$ 1.95 μm (2.15 μm), and its amplitude decreases to ~80% after annealing. Our experimental results reveal that, irrespective of the value of $d$, the spectral shifting occurs only during the first few annealing cycles and then remains stable for the following heating cycles.

The observed redshift may result from the observed morphology change of the tungsten nanodisks during thermal annealing. As can be seen in the high-magnification SEM images (see Supplementary Figure S4), the tungsten nanodisks seem to undergo a subtle structural reconstruction induced by the annealing to reach a new thermodynamically stable phase and eventually become more oblate-like spheroids after 10 heating cycles, which could change the resonance strength and dispersion of the nanodisk array, and consequently shift the cavity resonance responsible for metasurface absorption peak towards longer wavelengths[50]. The surfaces of the tungsten metasurface emitters also appear much rougher after annealing, leading to higher scattering losses and thus lower emittance. We note that, despite the slightly decreased emittance after annealing, the redshifted resonant peak for the $d$ = 360 nm metasurface emitter not only remains within the high EQE region, but it is also well aligned to the 1200 °C blackbody spectrum (shaded background) and the band-edge of the InGaAsSb PV cell. The below-bandgap thermal photons could, in principle, be further recycled with the aid of an optical filter[13].

Similar heating cycles were applied to the W-metasurface solar absorbers. As shown by SEM images in Figures 3a and 3b, the structural integrity of both uncoated and $HfO_2$-coated absorbers is well maintained after 10 heating cycles, and no discernible structural degradation is observed. In Figures 3d and 3e, we compare their spectral absorptance before and after 5 and 10 heating cycles, revealing that the absorptance after repeated heating cycles remains mostly unchanged as compared to its value before annealing, and stays constant after the first heating cycle. At short wavelengths, the absorptance is slightly enhanced after annealing, which is probably due to the slightly smaller and rounder W-nanodisks (see Supplementary Figure S5). However, it is unclear why the morphology change observed in metasurface emitters was not observed for our uncoated W-metasurface absorbers after the same number of annealing cycles (see insets to Figure 2d), although it could be inferred that the morphology change of nanodisks induced by thermal annealing may have been largely dependent upon the nanodisk size[38]. We further annealed the $HfO_2$-coated W-metasurface absorbers at 1200 °C for a total of 10 hours to investigate their long-term thermal stability, and the results shown in Figures 3c and 3f clearly reveal that both their structural and optical properties are well maintained after the prolonged heating, further demonstrating the superior thermal stability of the W-based metasurface solar absorbers.

For STPV application the spectral characteristics of the W-metasufaces should be retained at high operational temperatures. We characterized *in-situ* the absorptance spectra of the W-metasurface solar absorbers at 1200 °C. As shown in Figure 4 the solid curves, the absorption characteristics of both uncoated and $HfO_2$ coated absorbers are clearly unchanged at the elevated temperature of 1200 °C, except for the slightly increased absorptance at wavelengths $\lambda > 1.6$ μm, as compared to those absorptance spectra measured at room temperature shown by the dashed curves. We also notice that the absorptance of both absorbers in the infrared region seems to slightly increase with temperature. These could be due to the increased surface roughness and the unavoidable increased absorption of tungsten at high temperatures[51]. It is worth noting that the maximum temperature in our experiments is limited to the melting point of the silicon substrate. Therefore, we believe that our W-based metasurface absorbers and emitters, especially the $HfO_2$-coated ones, are likely to withstand operational temperatures much higher than 1200 °C when fabricated on a refractory substrate.

**Discussion**

The demonstrated W-metasurface absorbers and emitters are promising for further enhancing the current STPV efficiency due to their superior spectral selectivity and thermal stability. Compared to numerous absorbers and emitters reported elsewhere[20, 30, 31, 33, 36-39], the demonstrated spectral selectivity and thermal stability of our W-metasurfaces are unprecedented. Their optical and structural properties could potentially be further improved by using higher quality tungsten deposited by chemical vapor deposition (CVD) or ALD. Other metals such as tantalum or molybdenum are also suitable for high-performance refractory metasurface absorbers and emitters. In particular, tantalum has lower emittance in the near- and mid-infrared compared to tungsten[52], which could lead to solar absorbers with an even sharper cutoff and thermal emitters with less below-bandgap emission. Regardless of materials, our metasurface structures can be easily scaled up to a large surface area by using advanced fabrication techniques, such as deep UV photolithography or nanoimprint. Due to their compact size, they could potentially be integrated into more complex STPV systems with non-planar geometries[49].

We estimated the performance of a potential STPV system employing a fully integrated W-metasurface absorber/emitter intermediate structure based on a detailed balance calculation (see Methods). In the calculation, the after-annealing experimental data of the $HfO_2$-coated W-metasurface absorber and the $d = 360$ nm thermal emitter were used. In Figures 5a and 5b, we show the equilibrium temperature $T_{IS}$ and efficiency $\eta_{IS}$ of the intermediate structure as a function of solar concentration $N_s$ and emitter-to-absorber area ratio $A_e/A_a$. As can be seen, the intermediate structure can reach a high equilibrium temperature necessary for STPV over a wide range of operating conditions. For example, an intermediate structure with area ratio of 10 could reach $T_{IS} = 1200$ °C under solar concentration around $N_s = 2000$ suns. In Figure 5b, we find that the intermediate structure could achieve an efficiency higher than 70% when the area ratio is larger than 10 regardless of the solar concentration. For a given solar concentration, $\eta_{IS}$ increases with the emitter-to-absorber area ratio up to $A_e/A_a \sim 20$. Further increasing the area ratio has a much less effect on $\eta_{IS}$, because although the emitted power grows with $A_e$, the equilibrium temperature decreases with $A_e/A_a$ (see Figure 5a). Also note that the intermediate structure efficiency does not have a strong dependence on $N_s$: although higher solar concentration could result in more emitted power from the emitter due to higher equilibrium temperatures, at the same time it also increases the re-emission losses from the absorber. Finally, the projected full STPV efficiency $\eta_{STPV}$ for an InGaAsSb PV cell is illustrated in Figure 5c. It is clear that the STPV system could attain efficiency much higher than that of a stand-alone InGaAsSb PV cell (the dashed line represents the efficiency of the InGaAsSb cell, ~4% under direct solar illumination). The maximum STPV efficiency is as high as 18% at the equilibrium temperature of 1380 °C under solar concentration of 4000 suns. For comparison, the efficiency of the InGaAsSb PV cell is only 7.9% under the same solar concentration. More importantly, we find that $\eta_{STPV}$ could exceed that of the InGaAsSb cell at solar concentrations of as low as ~ 70 suns for $A_e/A_a$ = 5, and readily reach 10% at ~200 and 400 suns for two practical emitter-to-absorber area ratios $A_e/A_a$ = 5 and 15, respectively (see Figure 5d). These findings suggest that the potential STPV system employing our W-metasurface intermediate structure is indeed likely to outperform the recent STPV demonstrations based on other absorber/emitter intermediate structures, without resorting to very harsh operating conditions.

CONCLUSION

In conclusion, we have demonstrated tungsten-based refractory metasurfaces with desired spectral selectivity for STPV applications. The metasurface solar absorbers exhibit high absorption at visible and near-infrared wavelengths where the solar spectral intensity is significant, while the emittance is greatly suppressed at longer wavelengths to reduce the thermal radiative loss. The metasurface thermal emitters allow for wavelength selective emission matching the bandgap of narrow-gap photovoltaic cells and the spectral peak of blackbody radiation to maximize the thermal energy harvesting. We find the metasurface structural integrity and spectral response do not show any significant degradation following prolonged high temperature operation and repeated heating cycles between room temperature and up to 1200 °C. When the metasurface solar absorber and thermal emitter are fully integrated to form the intermediate structure, through a detailed balance calculation we estimated the overall STPV efficiency as high as 18.5% with a practical solar concentration and emitter-to-absorber area ratio, much higher than those achievable by stand-alone PV cells, therefore providing a path forward for high-performance STPV systems based on refractory metasurface structures.

METHODS

***Fabrication of tungsten metasurface absorbers/emitters.*** All metasurfaces were fabricated on 1 cm × 1 cm, double-sided polished, high-resistivity ($\rho > 10{,}000$ Ω-cm) silicon substrate, with an active area typically 1 mm × 1 mm. A 150-nm-thick W-film was first deposited on the substrate surface by DC sputtering at room temperature (power = 100 W, pressure = 3 mT), followed by an $Al_2O_3$ thin film using atomic layer deposition (ALD) at 250 °C using trimethylaluminum (TMA) and water precursors. After another 50-nm-thick W-film also by sputtering, a thin layer of chromium (Cr) was deposited by e-beam evaporation to be used as an etching mask for the nanodisks, which were defined by e-beam lithography (JEOL JBX-6300FS) using negative e-beam resist (Dow Corning XR-1541). The nanodisk patterns were transferred into the underlying Cr layer by inductively coupled plasma (ICP) etching with $Cl_2$/Ar chemistry, and subsequently into the W-film by fluorine-based reactive-ion etching (RIE). The Cr etching mask was removed thereafter by a wet chemical etching process, leaving behind square arrays of W-nanodisks on the $Al_2O_3$/W layers. $HfO_2$ protective coating was deposited by ALD at 250 °C using tetrakis (dimethylamido) hafnium (TDMAH) and water precursors.

***Room-temperature reflection measurements***. The room-temperature reflectance (angle of incidence of 20°) of both thin films and all of the fabricated metasurfaces were characterized using a variable angle spectroscopic ellipsometry (VASE) system (J. A. Woollam Co.) at wavelengths from 300 nm to 2500 nm. The unpolarized reflectance was obtained by averaging the p- and s-polarized reflectance. For angle dependent reflection measurements, the angle of incidence was varied from 20° to 60° with an increment of 10°. The complex dielectric functions of the W-films were extracted by fitting to the experimental data based on an air/tungsten/silicon three-layer structure using a Drude-Lorentz model.

***High-temperature reflection measurements and thermal annealing***. The *in-situ* high-temperature reflectance was characterized using a home-built measurement setup (see Supplementary Figure S3), which mainly comprises a high-temperature vacuum chamber (HeatWave Labs, Inc.), a white light laser source, and a spectrometer/detector suite (Ocean Optics, Inc.). After loading the sample, the vacuum chamber was evacuated to a base pressure of $< 10^{-5}$ Torr. The sample was then heated from 20 °C to the targeted temperatures controlled and monitored by a temperature controller (HeatWave Labs, Inc.). At each temperature, the reflection signal (angle of incidence of 12°) from the sample was collected and normalized to that from a silver mirror to extract its spectral reflectance. This setup was also used for thermal annealing of the samples. After reaching the base pressure, the sample was heated from room temperature to 1200 °C at a ramp rate of 10 °C/min. For multiple heating cycles, the sample was cooled back down to 20 °C at the same ramp rate to accomplish one heating cycle, and the same heating and cooling steps were repeated until the total number of heating cycles was achieved.

***XRD and SEM characterization***. X-ray diffraction (XRD) analyses of tungsten thin films were conducted using a Rigaku Ultima III diffractometer with a Cu Kα (1.5406 Å) X-ray source. Scanning electron microscopy (SEM) images of tungsten thin films and the fabricated W-metasurfaces were taken on a high-resolution FEI Nova NanoSEM 450 system.

***FDTD simulations***. Full-wave numerical simulations of metasurface absorbers and emitters were carried out in the frequency domain with periodic boundary conditions using commercial simulation software packages (COMSOL Multiphysics and CST Microwave Studio). The measured frequency dependent dielectric properties of tungsten and aluminum oxide were used in the simulations of the specular scattering S parameters, including the transmission $S_{21}$ and reflection $S_{11}$ coefficients, and the absorptance was then derived using $A = 1 - |S_{11}|^2$, as $S_{21}$ is zero due to the thick W-ground plane. The absorptance was also calculated by integrating the dissipated power density over the entire volume of the metasurface unit cell. No difference with respect to $1 - |S_{11}|^2$ was observed and, hence, high order diffraction modes are not excited in our structures. Although the absorber does not present continuous rotational invariance, it was numerically checked that its absorptance depends weakly on the azimuthal angle $\varphi$.

***STPV efficiency***. The equilibrium temperature $T_{\mathrm{IS}}$ of the intermediate structure was determined by the detailed balance equation $A_{\mathrm{a}}[P_{\mathrm{s}}(N_{\mathrm{s}}, T_{\mathrm{s}}) - P_{\mathrm{a}}(T_{\mathrm{IS}})] = A_{\mathrm{e}} P_{\mathrm{e}}(T_{\mathrm{IS}})$, where $P_{\mathrm{s}} = \int_0^{2\pi} d\varphi \int_0^{\theta_c} d\theta \sin\theta \cos\theta \int_0^{\infty} d\lambda\, A(\lambda, \theta, \varphi) I_{\mathrm{BB}}(\lambda, T_{\mathrm{s}})$ is the power absorbed, $P_{\mathrm{a}} = \int_0^{2\pi} d\varphi \int_0^{\pi/2} d\theta \sin\theta \cos\theta \int_0^{\infty} d\lambda\, A(\lambda, \theta, \varphi) I_{\mathrm{BB}}(\lambda, T_{\mathrm{IS}})$ is the re-emission loss, and $P_{\mathrm{e}} = \int_0^{2\pi} d\varphi \int_0^{\pi/2} d\theta \sin\theta \cos\theta \int_0^{\infty} d\lambda\, E(\lambda, \theta, \varphi) I_{\mathrm{BB}}(\lambda, T_{\mathrm{IS}})$ is the power emitted towards the PV cell. Here, $I_{\mathrm{BB}}(\lambda, T)$ is the blackbody spectral irradiance at temperature $T$, $T_{\mathrm{s}}$ is the temperature of the sun, $N_s$ is the number of solar concentration, and $\theta_c = \sin^{-1}\sqrt{\frac{N_s \Omega_s}{\pi}}$ with $\Omega_s = 68.5$ µSr represents the solid angle subtended by the sun. We neglected the thermal radiation from the solar cell to the emitter. The efficiency of the intermediate structure is $\eta_{IS} = \frac{A_e P_e}{A_a P_0}$, where $P_0$ is the incident power on the absorber (given by $P_s$ with $A = 1$). The STPV efficiency for a solar cell at temperature $T_{PV}$ is $\eta_{STPV} = \eta_{IS} \times U \times v \times m$, where $U(T_{IS})$ is the ultimate efficiency, $v(T_{IS}, T_{PV})$ is the open circuit factor, and $m(T_{IS}, T_{PV})$ is the impedance matching factor, all computed using the measured external quantum efficiency of the PV cell[10].

## ASSOCIATED CONTENT

**Supporting Information**

Figures S1-S5.

## AUTHOR INFORMATION

**Corresponding Author**

*E-mail: chenht@lanl.gov

Phone: 1-505-665-7365

Center for Integrated Nanotechnologies, Los Alamos National Laboratory

MS K771, PO Box 1663, Los Alamos, New Mexico 87545, USA

**Author Contributions**

A.K.A., A.J.T., D.A.R.D., M.S. and H.T.C. conceived the project. W.K.K. and C.C.C. performed numerical simulations and designed the metasurface structures. C.C.C. planned and led all the experiments for the project. J.N. assisted the metasurface sample fabrication. T.S.L. assisted the characterization of the reflectance at room temperature. M.S. built the high-temperature characterization setup and, together with C.C.C. and A.K.A, performed the *in-situ* reflectance measurements at high temperature. W.K.K. and D.A.R.D. estimated the performance of the potential STPV system. C.C.C. and H.T.C. wrote the manuscript with major contributions from W.K.K. and D.A.R.D., and all authors discussed the results.

**Notes**

The authors declare no competing financial interest.

ACKNOWLEDGMENT

The authors thank discussions with Eli Ben-Naim, Stuart Trugman, and Nikolai Sinitsyn. C.C.C. gratefully acknowledges the CINT Integration Lab staff, Anthony R. James, Willard Ross, Denise Webb, Joseph Lucero, and Doug Pete for their assistance in device fabrication and valuable discussions. The authors acknowledge support from the Los Alamos National Laboratory LDRD program. W.K.K. acknowledges the Center for Nonlinear Studies (CNLS) for partial support. This work was performed, in part, at the Center for Integrated Nanotechnologies, a U.S. Department of Energy, Office of Basic Energy Sciences Nanoscale Science Research Center operated jointly by Los Alamos and Sandia National Laboratories. Los Alamos National Laboratory, an affirmative action/equal opportunity employer, is operated by Los Alamos National Security, LLC, for the National Nuclear Security Administration of the U.S. Department of Energy under Contract No. DE-AC52-06NA25396.

REFERENCES

1. Chapin, D. M.; Fuller, C. S.; Pearson, G. L. *J. Appl. Phys.* **1954,** 25(5), 676-677.

2. Shockley, W.; Queisser, H. J. *J. Appl. Phys.* **1961,** 32(3), 510-519.

3. Dimroth, F.; Grave, M.; Beutel, P.; Fiedeler, U.; Karcher, C.; Tibbits, T. N. D.; Oliva, E.; Siefer, G.; Schachtner, M.; Wekkeli, A.; Bett, A. W.; Krause, R.; Piccin, M.; Blanc, N.; Drazek, C.; Guiot, E.; Ghyselen, B.; Salvetat, T.; Tauzin, A.; Signamarcheix, T.; Dobrich, A.; Hannappel, T.; Schwarzburg, K. *Prog. Photovolt. Res. Appl.* **2014,** 22(3), 277-282.

4. Trupke, T.; Green, M. A.; Würfel, P. *J. Appl. Phys.* **2002,** 92(7), 4117-4122.

5. Trupke, T.; Green, M. A.; Würfel, P. *J. Appl. Phys.* **2002,** 92(3), 1668-1674.

6. Atwater, H. A.; Polman, A. *Nat. Mater.* **2010,** 9(3), 205-213.

7. Burschka, J.; Pellet, N.; Moon, S.-J.; Humphry-Baker, R.; Gao, P.; Nazeeruddin, M. K.; Gratzel, M. *Nature* **2013,** 499(7458), 316-319.

8. Spirkl, W.; Ries, H. *J. Appl. Phys.* **1985,** 57(9), 4409-4414.

9. Davies, P. A.; Luque, A. *Sol. Energy Mater. Sol. Cells* **1994,** 33(1), 11-22.

10. Rephaeli, E.; Fan, S. *Opt. Express* **2009,** 17(17), 15145-15159.

11. Harder, N.-P.; Würfel, P. *Semicond. Sci. Technol.* **2003,** 18(5), S151.

12. Lenert, A.; Bierman, D. M.; Nam, Y.; Chan, W. R.; Celanovic, I.; Soljacic, M.; Wang, E. N. *Nat. Nano.* **2014,** 9(2), 126-130.

13. Bierman, D. M.; Lenert, A.; Chan, W. R.; Bhatia, B.; Celanović, I.; Soljačić, M.; Wang, E. N. *Nat. Energy* **2016,** 1, 16068.

14. Rinnerbauer, V.; Lenert, A.; Bierman, D. M.; Yeng, Y. X.; Chan, W. R.; Geil, R. D.; Senkevich, J. J.; Joannopoulos, J. D.; Wang, E. N.; Soljačić, M.; Celanovic, I. *Adv. Energy Mater.* **2014,** 4(12), 1400334.

15. Shimizu, M.; Kohiyama, A.; Yugami, H. *J. Photonics Energy* **2015,** 5(1), 053099.

16. Kohiyama, A.; Shimizu, M.; Yugami, H. *Appl. Phys. Express* **2016,** 9(11), 112302.

17. Lampert, C. M. *Sol. Energy Mater.* **1979,** 1(5), 319-341.

18. Narayanaswamy, A.; Chen, G. *Phys. Rev. B* **2004,** 70(12), 125101.

19. Li, X.-F.; Chen, Y.-R.; Miao, J.; Zhou, P.; Zheng, Y.-X.; Chen, L.-Y.; Lee, Y.-P. *Opt. Express* **2007,** 15(4), 1907-1912.

20. Khodasevych, I. E.; Wang, L.; Mitchell, A.; Rosengarten, G. *Adv. Opt. Mater.* **2015,** 3(7), 852-881.

21. Azad, A. K.; Kort-Kamp, W. J. M.; Sykora, M.; Weisse-Bernstein, N. R.; Luk, T. S.; Taylor, A. J.; Dalvit, D. A. R.; Chen, H.-T. *Sci. Rep.* **2016,** 6, 20347.

22. Li, W.; Guler, U.; Kinsey, N.; Naik, G. V.; Boltasseva, A.; Guan, J.; Shalaev, V. M.; Kildishev, A. V. *Adv. Mater.* **2014,** 26(47), 7959-7965.

23. Wang, Y.; Zhou, L.; Zheng, Q.; Lu, H.; Gan, Q.; Yu, Z.; Zhu, J. *Appl. Phys. Lett.* **2017,** 110(20), 201108.

24. Chubb, D. L., *Fundamentals of thermophotovoltaic energy conversion*. Elsevier: 2007.

25. Bett, A. W.; Sulima, O. V. *Semicond. Sci. Technol.* **2003,** 18(5), S184.

26. Dashiell, M. W.; Beausang, S. F.; Ehsani, H.; Nichols, G. J.; Depoy, D. M.; Danielson, L. R.; Talamo, P.; Rahner, K. D.; Brown, E. J.; Burger, S. R.; Fourspring, P. M.; Topper, W. F.; Baldasaro, P. F.; Wang, C. A.; Huang, R. K.; Connors, M. K.; Turner, G. W.; Schellenbarger, Z. A.; Taylor, G.; Li, J.; Martinelli, R.; Donetski, D.; Anikeev, S.; Belenky, G. L.; Luryi, S. *IEEE Trans. Electron Dev.* **2006,** 53(12), 2879.

27. Liu, J.; Guler, U.; Lagutchev, A.; Kildishev, A.; Malis, O.; Boltasseva, A.; Shalaev, V. M. *Opt. Mater. Express* **2015,** 5(12), 2721-2728.

28. Mason, J. A.; Adams, D. C.; Johnson, Z.; Smith, S.; Davis, A. W.; Wasserman, D. *Opt. Express* **2010,** 18(24), 25192-25198.

29. Park, J. H.; Han, S. E.; Nagpal, P.; Norris, D. J. *ACS Photon.* **2016,** 3(3), 494-500.

30. Rinnerbauer, V.; Yeng, Y. X.; Chan, W. R.; Senkevich, J. J.; Joannopoulos, J. D.; Soljačić, M.; Celanovic, I. *Opt. Express* **2013,** 21(9), 11482-11491.

31. Lin, S. Y.; Moreno, J.; Fleming, J. G. *Appl. Phys. Lett.* **2003,** 83(2), 380-382.

32. Nagpal, P.; Han, S. E.; Stein, A.; Norris, D. J. *Nano Lett.* **2008,** 8(10), 3238-3243.

33. Arpin, K. A.; Losego, M. D.; Cloud, A. N.; Ning, H.; Mallek, J.; Sergeant, N. P.; Zhu, L.; Yu, Z.; Kalanyan, B.; Parsons, G. N.; Girolami, G. S.; Abelson, J. R.; Fan, S.; Braun, P. V. *Nat. Commun.* **2013,** 4, 2630.

34. Wang, Z.; Luk, T. S.; Tan, Y.; Ji, D.; Zhou, M.; Gan, Q.; Yu, Z. *Appl. Phys. Lett.* **2015,** 106(10), 101104.

35. Liu, X.; Tyler, T.; Starr, T.; Starr, A. F.; Jokerst, N. M.; Padilla, W. J. *Phys. Rev. Lett.* **2011,** 107(4), 045901.

36. Shemelya, C.; DeMeo, D.; Latham, N. P.; Wu, X.; Bingham, C.; Padilla, W.; Vandervelde, T. E. *Appl. Phys. Lett.* **2014,** 104(20), 201113.

37. Woolf, D.; Hensley, J.; Cederberg, J. G.; Bethke, D. T.; Grine, A. D.; Shaner, E. A. *Appl. Phys. Lett.* **2014,** 105(8), 081110.

38. Yokoyama, T.; Dao, T. D.; Chen, K.; Ishii, S.; Sugavaneshwar, R. P.; Kitajima, M.; Nagao, T. *Adv. Opt. Mater.* **2016,** 4(12), 1987-1992.

39. Hasan, D.; Pitchappa, P.; Wang, J.; Wang, T.; Yang, B.; Ho, C. P.; Lee, C. *ACS Photon.* **2017,** 4(2), 302-315.

40. Dyachenko, P. N.; Molesky, S.; Petrov, A. Y.; Störmer, M.; Krekeler, T.; Lang, S.; Ritter, M.; Jacob, Z.; Eich, M. *Nat. Commun.* **2016,** 7, 11809.

41. Stelmakh, V.; Rinnerbauer, V.; Geil, R. D.; Aimone, P. R.; Senkevich, J. J.; Joannopoulos, J. D.; Soljačić, M.; Celanovic, I. *Appl. Phys. Lett.* **2013,** 103(12), 123903.

42. Landy, N. I.; Sajuyigbe, S.; Mock, J. J.; Smith, D. R.; Padilla, W. J. *Phys. Rev. Lett.* **2008,** 100(20), 207402.

43. Tao, H.; Bingham, C. M.; Strikwerda, A. C.; Pilon, D.; Shrekenhamer, D.; Landy, N. I.; Fan, K.; Zhang, X.; Padilla, W. J.; Averitt, R. D. *Phys. Rev. B* **2008,** 78(24), 241103.

44. Liu, N.; Mesch, M.; Weiss, T.; Hentschel, M.; Giessen, H. *Nano Lett.* **2010,** 10(7), 2342-2348.

45. Arpin, K. A.; Losego, M. D.; Braun, P. V. *Chem. Mater.* **2011,** 23(21), 4783-4788.

46. Chan, W.; Huang, R.; Wang, C.; Kassakian, J.; Joannopoulos, J.; Celanovic, I. *Sol. Energy Mater. Sol. Cells* **2010,** 94(3), 509-514.

47. Yeng, Y. X.; Chan, W. R.; Rinnerbauer, V.; Joannopoulos, J. D.; Soljačić, M.; Celanovic, I. *Opt. Express* **2013,** 21(S6), A1035-A1051.

48. Watts, C. M.; Liu, X. L.; Padilla, W. J. *Adv. Mater.* **2012,** 24(23), Op98-Op120.

49. Wu, C.; Neuner, B.; John, J.; Milder, A.; Zollars, B.; Savoy, S.; Shvets, G. *J. Opt.* **2012,** 14(2), 024005.

50. Chen, H.-T. *Opt. Express* **2012,** 20(7), 7165-7172.

51. Yeng, Y. X.; Ghebrebrhan, M.; Bermel, P.; Chan, W. R.; Joannopoulos, J. D.; Soljačić, M.; Celanovic, I. *Proc. Natl. Acad. Sci. U.S.A*. **2012,** 109(7), 2280-2285.

52. Palik, E. D., *Handbook of optical constants of solids*. Academic Press: 1985.

FIGURES

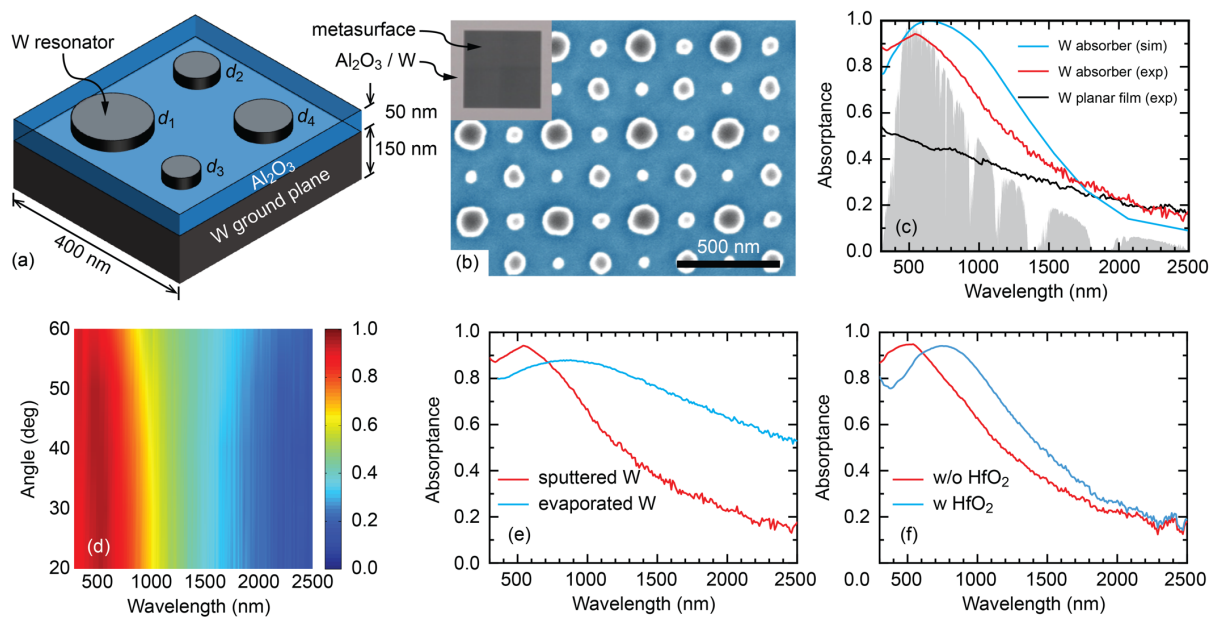

**Figure 1.** Tungsten metasurface solar absorbers. (a) Unit cell schematic of the absorber structure consisting of four W-nanodisks with optimized diameters $d_1$ = 140 nm, $d_2$ = 80 nm, $d_3$ = 60 nm, $d_4$ = 100 nm, $Al_2O_3$ spacer, and bottom W-ground plane. (b) Top-view false colored SEM image of the fabricated W-metasurface solar absorber. Inset: an optical image of the fabricated sample. (c) Measured (red) and simulated (cyan) absorptance for W-metasurface absorbers, and measured absorptance for a plain W-film (black). Shaded background: AM 1.5 solar spectrum. (d) Measured absorptance as a function of angle of incidence and wavelength. (e) A comparison of absorptance spectra between metasurfaces using sputtered (red) and evaporated (cyan) W-films. (f) A comparison of absorptance spectra between metasurfaces with (cyan) and without (red) a 20-nm-thick $HfO_2$ protective layer.

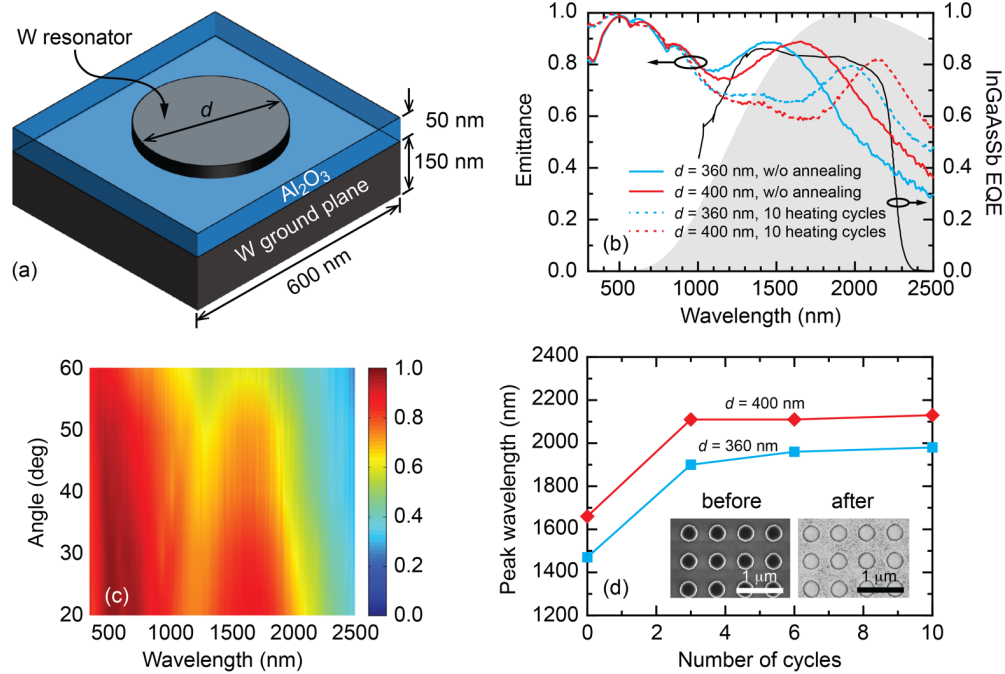

**Figure 2.** Tungsten metasurface thermal emitters. (a) Unit cell schematic of the metasurface thermal emitter consisting of a single W-nanodisk, $Al_2O_3$ spacer, and bottom W-ground plane. (b) Emittance for two selected W-nanodisk diameters before annealing (solid) and after 10 heating cycles (dotted). The black curve is for the external quantum efficiency (EQE) of the InGaAsSb PV cell, and the shaded background indicates the blackbody spectrum at 1200 °C. (c) Angle dependence of emittance for the $d$ = 400 nm metasurface emitter. (d) Resonance peak wavelength of the metasurface emitters as a function of number of heating cycles. Insets: SEM images of the metasurface emitters before (left) and after 10 heating cycles.

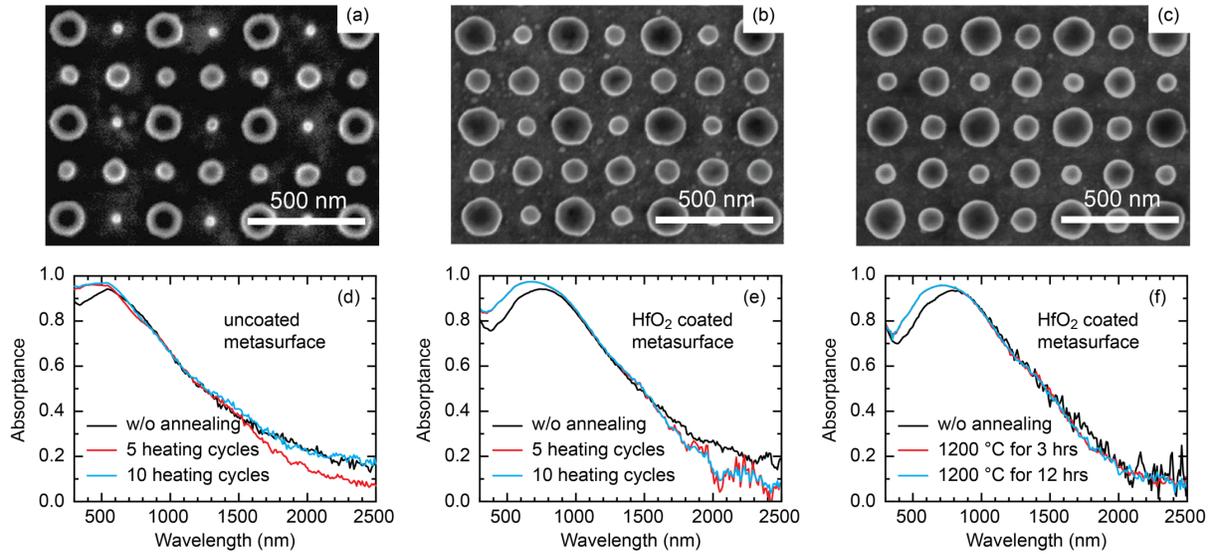

**Figure 3.** Thermal stability of tungsten metasurface solar absorbers. The top row shows SEM images for: (a) metasurface without protective coating after 10 heating cycles, (b) $HfO_2$-coated metasurface after 10 heating cycles, and (c) $HfO_2$-coated metasurface after annealing at 1200 °C for 10 hours. The bottom row plots the absorptance spectra measured at room temperature for: (d) metasurface without protective coating before annealing (black) and after 5 (red) and 10 (cyan) heating cycles, (e) same as in (d) but for $HfO_2$-coated metasurface, and (f) $HfO_2$-coated metasurface before (black) and after annealing at 1200 °C for 3 (red) and 10 (cyan) hours.

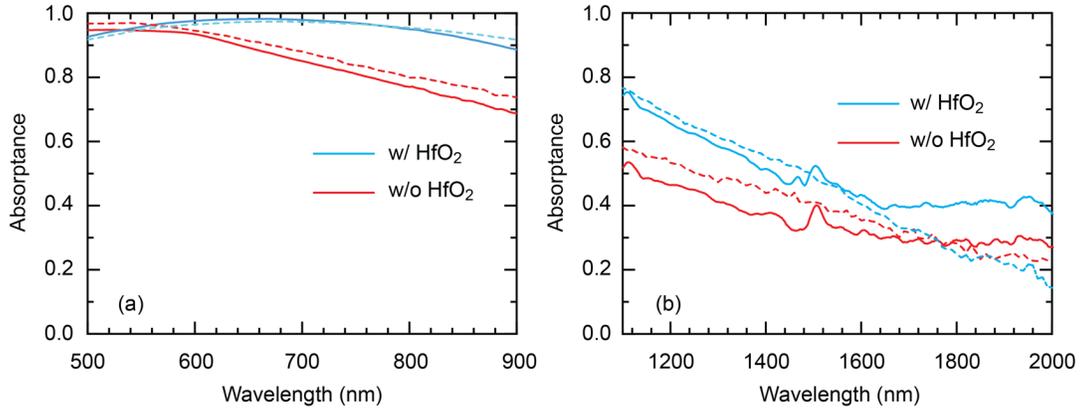

**Figure 4.** Absorptance spectra measured *in-situ* at 1200 °C for tungsten metasurface solar absorbers with (cyan solid) and without (red solid) the $HfO_2$ protective coating. Left and right panels are measured separately using different detectors to cover wavelengths up to 2000 nm. The dashed curves are for the corresponding samples after 10 heating cycles and measured at room temperature, replot from Figures 3d and 3e.

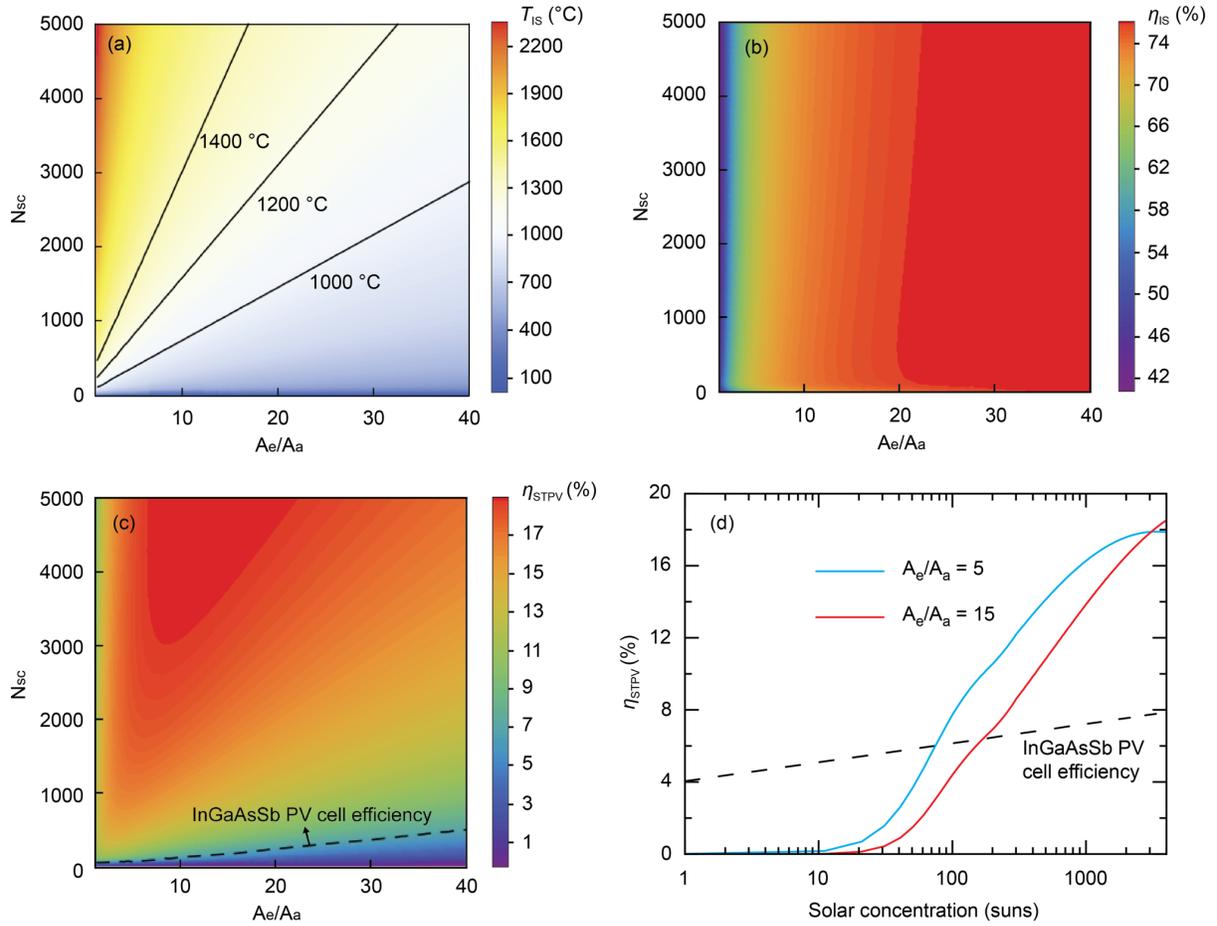

**Figure 5.** STPV efficiency with the tungsten metasurface intermediate structure. (a) Calculated equilibrium temperature of the intermediate structure, (b) intermediate structure efficiency and (c) STPV efficiency as a function of solar concentration and emitter-to-absorber area ratio. In (c) the black dashed curve corresponds to the InGaAsSb PV cell efficiency (~4%) under direct solar illumination. (d) STPV efficiency as a function of solar concentration for area ratios equal to 5 (cyan) and 15 (red). The black dashed curve is the InGaAsSb PV cell efficiency as a function of solar concentration.

# High-Temperature Refractory Metasurfaces for Solar Thermophotovoltaic Energy Harvesting

Chun-Chieh Chang[†], Wilton J. M. Kort-Kamp[‡,§], John Nogan[∥], Ting S. Luk[∥], Abul K. Azad[†], Antoinette J. Taylor[⊥], Diego A. R. Dalvit[§], Milan Sykora[#], and Hou-Tong Chen[*,†]

[†]Center for Integrated Nanotechnologies, Los Alamos National Laboratory, Los Alamos, New Mexico 87545, USA

[‡]Center for Nonlinear Studies, Los Alamos National Laboratory, Los Alamos, New Mexico 87545, USA

[§]Theoretical Division, Los Alamos National Laboratory, Los Alamos, New Mexico 87545, USA

[∥]Center for Integrated Nanotechnologies, Sandia National Laboratories, Albuquerque, New Mexico, 87123, USA

[⊥]Chemistry, Life, and Earth Sciences Directorate, Los Alamos National Laboratory, Los Alamos, New Mexico 87545, USA

[#]Chemistry Division, Los Alamos National Laboratory, Los Alamos, New Mexico 87545, USA

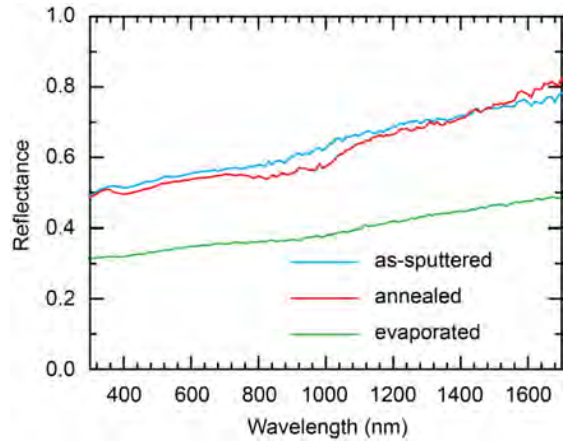

**Figure S1.** Reflectance spectra of as-sputtered, annealed at 850 °C, and evaporated tungsten films.

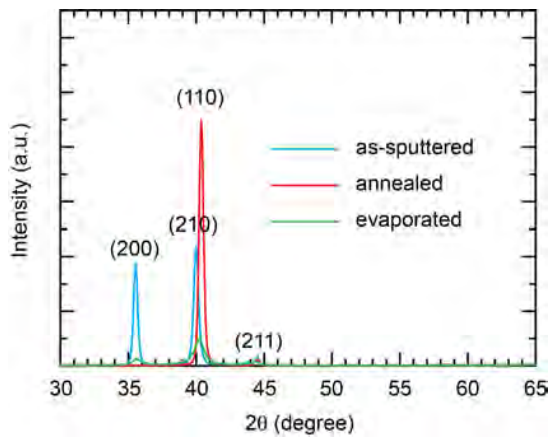

**Figure S2.** Structural properties of tungsten thin films. X-ray diffraction spectra of sputtered tungsten films before and after annealing at 850 °C and evaporated tungsten films. The as-deposited sputtered tungsten films are amorphous and show (200), (210), and (211) W-β phase peaks, whereas the annealed sputtered films are more crystalline and exhibit a predominant (110) W-α phase peak. The evaporated tungsten films show a much poorer crystalline order compared to the sputtered tungsten films and are in the β-phase[1, 2].

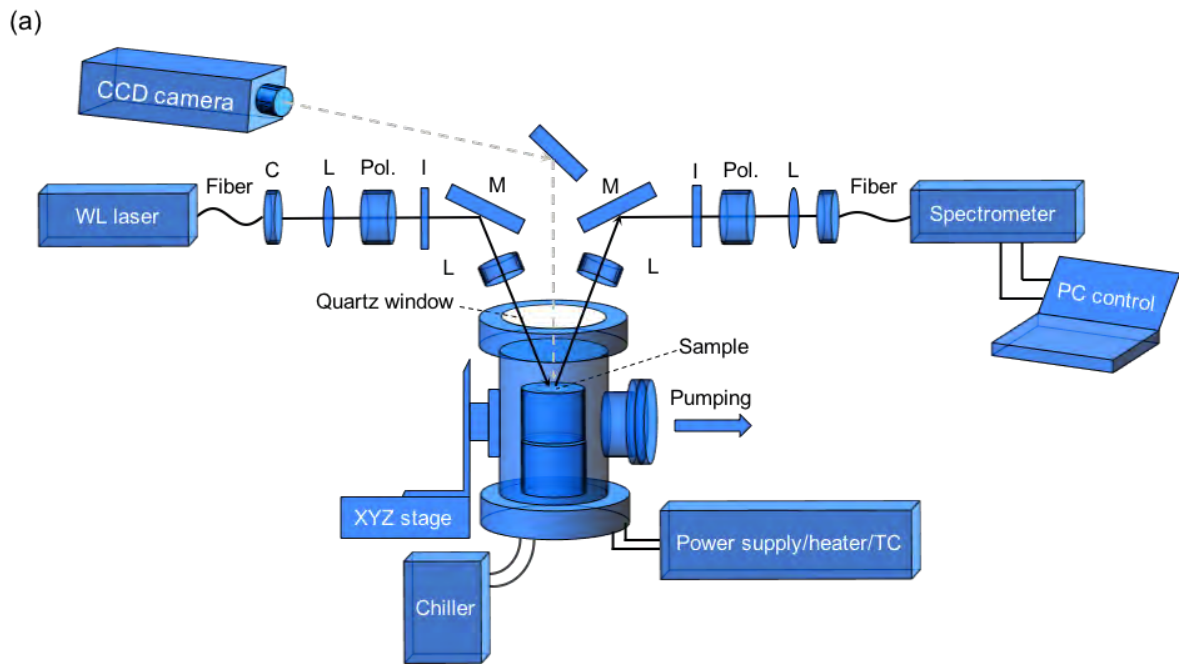

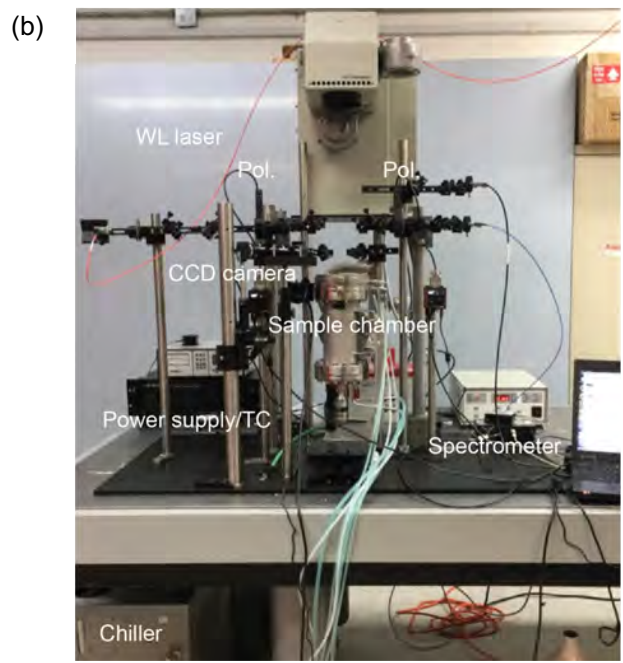

**Figure S3.** High-temperature characterization set-up. (a) The schematic and (b) photograph of the set-up. Acronyms in (a) and (b): WL: white light, C: collimator, L: lens, Pol.: polarizer, I: iris, M: mirror, PC: personal computer, TC: temperature control.

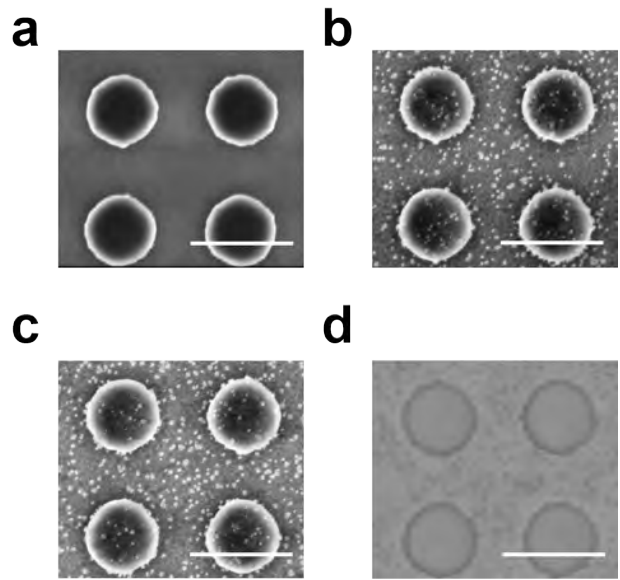

**Figure S4.** SEM images of the tungsten metamaterial emitters (a) before and after (b) 3, (c) 6, and (d) 10 heating cycles between room temperature and 1200 °C. Scale bars are 500 nm.

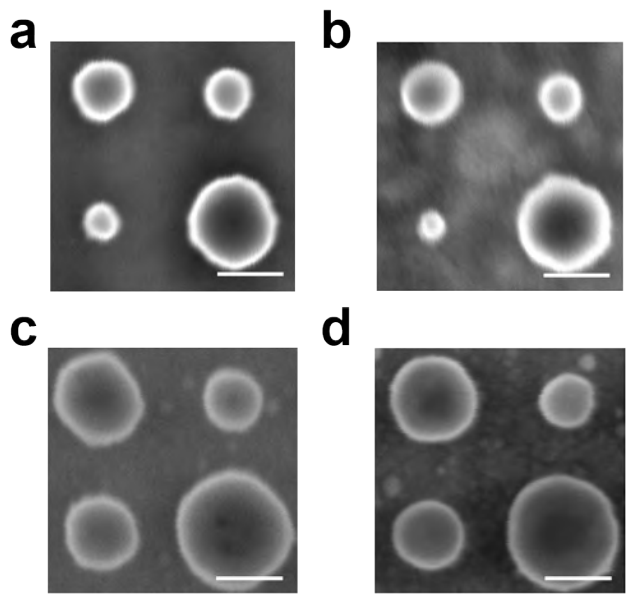

**Figure S5.** SEM images of a unit cell of tungsten metasurface absorbers (a) before and (b) after 10 heating cycles between room temperature and 1200 °C. (c) Same as (a), but for tungsten absorbers coated with an HfO$_2$ protective layer. (d) Same as (b), but for HfO$_2$-coated absorbers. Scale bars are 100 nm.

1. Vink, T. J.; Walrave, W.; Daams, J. L. C.; Dirks, A. G.; Somers, M. A. J.; van den Aker, K. J. A. *J. Appl. Phys.* **1993,** 74(2), 988-995.

2. Stelmakh, V.; Rinnerbauer, V.; Joannopoulos, J. D.; Soljačić, M.; Celanovic, I.; Senkevich, J. J.; Tucker, C.; Ives, T.; Shrader, R. *J. Vac. Sci. Technol., A* **2013,** 31(6), 061505.